\documentclass[aps,prb,amsmath,amssymb,notitlepage,showkeys,twocolumn,superscriptaddress]{revtex4-1}

\usepackage[]{graphicx}
\usepackage{subfigure}
\usepackage{float}
\usepackage{dcolumn}
\usepackage{bm}
\usepackage[colorlinks=true,citecolor=blue,linkcolor=blue,urlcolor=blue,bookmarks=true]{hyperref}
\usepackage[sort&compress]{natbib}
\newcommand{\fref}[1]{Fig.~\ref{#1}}
\renewcommand{\eqref}[1]{Eq.~(\ref{#1})}
\newcommand{\eref}[1]{(\ref{#1})}
\newcommand{\sfref}[2]{Fig.~\ref{#1}\hyperref[#1]{#2}}
\newcommand{\sref}[2]{\ref{#1}\hyperref[#1]{#2}}
\newcommand{\bto}{$\beta$-TeVO$_4$ }
\newcommand{\btos}{$\beta$-TeVO$_4$}
\begin{document}
%
%
\title{\texorpdfstring{Magnetic-field-induced reorientation in the SDW and the spin-stripe phases of the frustrated spin-$1/2$ chain compound \bto }{Magnetic-field-induced reorientation in the SDW and the spin-stripe phases of the frustrated spin-1/2 chain compound beta-TeVO4}}
\author{Mirta~Herak}
\email{mirta@ifs.hr}
\author{Nikolina Novosel}
\author{Martina~Dragi\v{c}evi\'{c}}
\affiliation{Institute of Physics, Bijeni\v{c}ka c.\ 46, HR-10000 Zagreb, Croatia}
\author{Thierry Guizouarn}
\author{Olivier Cador}
\affiliation{Institut des Sciences Chimiques de Rennes UMR 6226, Universit\'{e} de Rennes 1, Campus de Beaulieu, 35042 Rennes, France}
\author{Helmuth~Berger}
\affiliation{Institut de Physique de la Mati\`{e}re Complexe, EPFL, CH-1015 Lausanne, Switzerland}
\author{Matej~Pregelj}
\affiliation{Jo\v{z}ef Stefan Institute, Jamova 39, 1000 Ljubljana, Slovenia}
\author{Andrej Zorko}
\author{Denis Ar\v{c}on}
\affiliation{Jo\v{z}ef Stefan Institute, Jamova 39, 1000 Ljubljana, Slovenia}
\affiliation{Faculty of Mathematics and Physics, Jadranska 19, 1000 Ljubljana, Slovenia}
%
%
%
\begin{abstract}
 $\beta$-TeVO$_4$ is a frustrated spin 1/2 zig-zag chain system, where spin-density-wave (SDW), vector-chiral (VC) and an exotic dynamic spin-stripe phase compete at low temperatures. Here we use torque magnetometry to study the anisotropy of these phases in magnetic fields of up to 5~T. Our results show that the magnetic-field-induced spin reorientation occurs in the SDW and in the spin-stripe phases for $\mu_0 H\geq 2$~T. The observed spin reorientation is a new element of the anisotropic phase diagram for the field directions in the $ac$ and $a^*b$ crystallographic planes. The presented results should help establishing the model of anisotropic magnetic interactions, which are responsible for the formation of complex magnetic phases in \bto and similar quantum systems.
\end{abstract}
%
\maketitle
\section{Introduction}
\begin{figure}[tb]
	\centering
		\includegraphics[width=0.65\columnwidth]{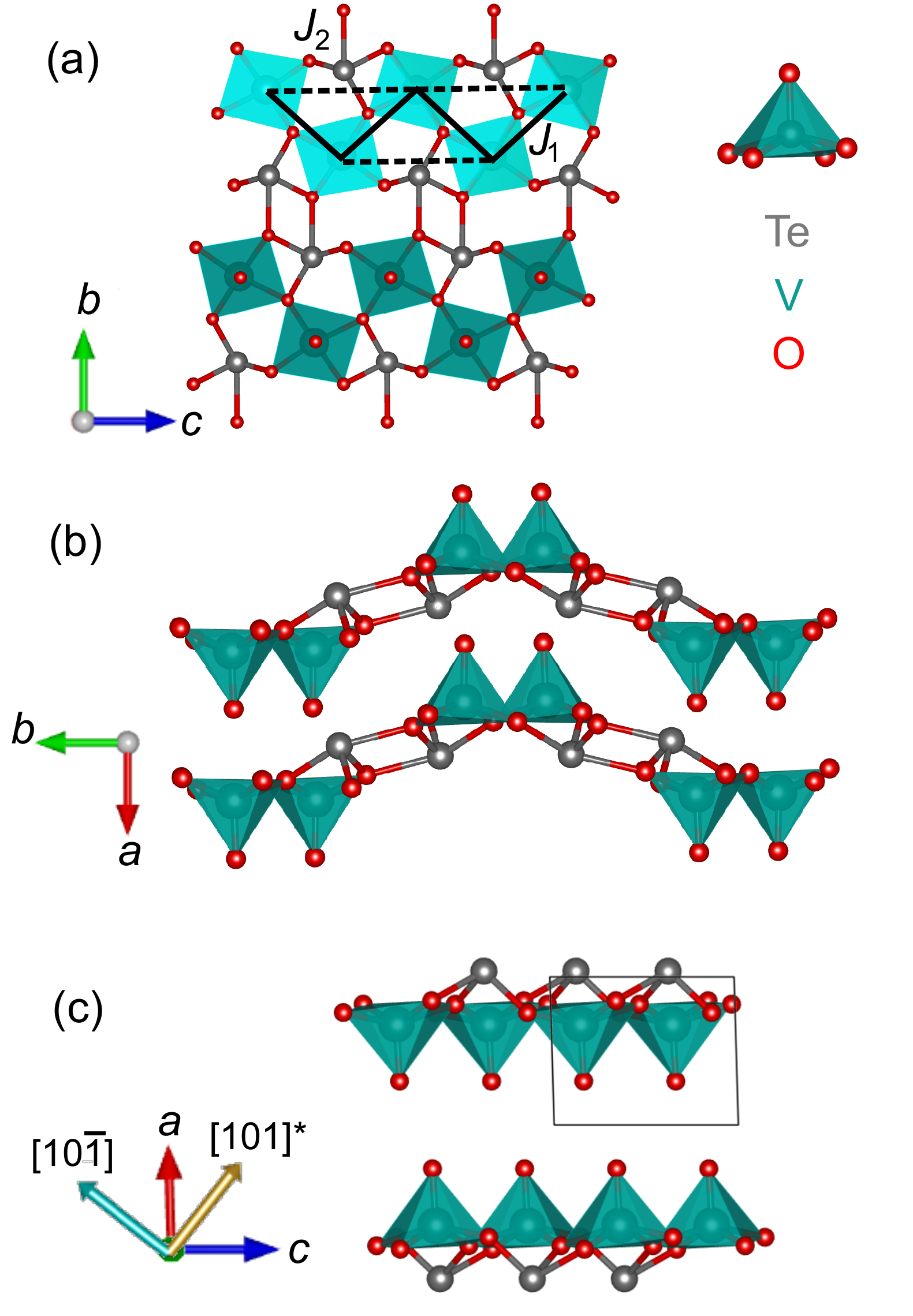}
	\caption{The crystal structure of $\beta$-TeVO$_4$. (a) $bc$ plane. The zig-zag chains run along the $c$ axis. (b)  $a^*b$ plane. (c) $ac$ plane. The experimentally determined magnetic eigenaxes at lowest temperature, $[1\:0\:\overline{1}]$ and $[1\:0\:1]^*$, are also plotted along with the crystal axes.}
	\label{fig1}
\end{figure}

\indent Frustrated low-dimensional spin systems often exhibit complex magnetic phases at low temperatures \cite{Lacroix-2011}, in contrast to simple collinear ferro- and antiferromagnets \cite{Blundell}. In frustrated spin-$1/2$ chains competition between the nearest-neighbor ferromagnetic (FM) and the next-nearest-neighbor antiferromagnetic (AFM) interactions leads to a number of different ground states as a function of magnetic field. For example, the spin multipolar phases, i.e.\ spin nematic phases \cite{Hikara-2008, Sudan-2009} are predicted to develop just below the saturation, which complies with the recent experimental studies of LiCuSbO$_4$ \cite{Grafe-2017,Bosiocic-2017} and LiCuVO$_4$ \cite{Orlova-2017,Buttgen-2014}. In weak magnetic fields, however, the presence of interchain interactions can stabilize the magnetically ordered spin-spiral states in which the spontaneous or induced electric polarization is observed \cite{Ruff-2019,Park-2007,Rusydi-2008,Mack-2017}, whereas the latter is absent at intermediate magnetic field, where collinear spin density wave (SDW) phase develops.\\
\indent One of the key ingredients that add to the complexity of the phase diagrams of these systems is the magnetic anisotropy which is responsible for arrangement of spins in the magnetically ordered phases of real materials and defines the details of phase boundaries between the competing phases in vanishing, as well as finite magnetic field. Applying a finite magnetic field can induce a phase transition between different phases, but it can also cause a reorientation of magnetic moments within a single phase. The field induced spin-flop in easy-axis antiferromagnets is the simplest example of such spin reorientation \cite{Neel-1952}, where the spin-flop field depends on the magnetic anisotropy in a particular antiferromagnet. The study of spin reorientations in antiferromagnets in finite magnetic fields thus represents a tool for probing their magnetic anisotropy.\\
\indent In this work we present a study of a frustrated spin-$1/2$ compound \btos, where an interplay of frustration and exchange anisotropies leads to complex antiferromagnetic phases that interchange as a function of magnetic field and temperature \cite{Pregelj-2016}. While the anisotropic phase diagram in this system has been studied in some detail \cite{Savina-2015,Weickert-2016,Pregelj-2019}, the possible occurrence of field-induced spin reorientation phenomena within the specific phases has not been explored so far.\\
\indent \bto crystallizes in the monoclinic space group $P2_1/c$ \cite{Meunier-1973,Pregelj-2015}. The spin-1/2 zig-zag chains are formed by the distorted VO$_5$ pyramids sharing their corners and running along the $c$ axis \cite{Meunier-1972, Savina-2011} (\fref{fig1}). The ferromagnetic nearest neighbor interaction of $J_1/k_B=-38$~K and the antiferromagnetic next-nearest neighbor interaction $J_2\approx -J_1$ between the V$^{4+}$ $S=1/2$ spins were determined experimentally \cite{Pregelj-2018}, whereas a lattice of weakly coupled frustrated spin chains was also suggested by theoretical considerations \cite{Saul-2014}. Weak interchain interactions are responsible for the three different magnetically long-range ordered phases in low magnetic fields ($\mu_0 H<2$~T), at temperatures $T_{N_1}$=4.65~K, $T_{N_2}$=3.28~K and $T_{N_3}$= 2.26~K \cite{Savina-2011}. These phases were characterized in a comprehensive study employing high-field magnetization, specific heat and neutron diffraction measurements \cite{Pregelj-2015}. The spin-density-wave (SDW) phase with the collinear incommensurate amplitude-modulated magnetic order ($T_{N_2} < T < T_{N_1}$) is followed by a dynamic spin-stripe phase ($T_{N_3}<T<T_{N_2}$) that still has a SDW character, and finally by a vector chiral (VC) phase, which has an elliptical-spiral magnetic order ($T < T_{N_3}$) \cite{Pregelj-2015}. Phase boundaries depend not only on the magnitude of the magnetic field, but also on the field direction \cite{Savina-2015, Weickert-2016,Pregelj-2019PRB}, which hints to an important role of magnetic anisotropy \cite{Singhania-2020}. Furthermore, the spin-stripe phase was recently proposed to host a new type of elementary excitation named ''wigglon'' which represents a bound state of two orthogonally polarized phason quasi-particles where higher-order magnetic anisotropies play a vital role \cite{Pregelj-2019}.\\
\indent In order to explore the magnetic anisotropy in the ordered states of $\beta$-TeVO$_4$, we studied the rearrangement of magnetic moments in applied magnetic field by employing torque magnetometry measurements. We report the full anisotropic $(T,\mathbf{H})$ phase diagram where the phase boundaries are determined as the magnetic field rotates in specific crystal planes. Moreover, our analysis of the torque curves shows that the field-induced spin reorientation occurs in the SDW and the spin-stripe phases, which we add as a new element to the anisotropic phase diagram of \btos.
\section{Experimental}\label{sec:exp}
\indent A high quality single crystal of monoclinic \bto with distinguished crystallographic planes $a^*b$, $bc$ and $ac$ was grown from TeO$_2$ and VO$_2$ powders by chemical vapour transport reaction, using two-zone furnace and TeCl$_4$ as a transport agent \cite{Pregelj-2015}. The same crystal was used in our previous publication \cite{Pregelj-2016}.\\
\indent The magnetic torque in the applied magnetic field $\mu_0 H=1$ and 2~T was measured by a highly sensitive custom-made torque magnetometer based on the torsion of a thin quartz fibre. The magnetic field was supplied by the Cryogenic Consultants 5-T split-coil superconducting magnet with a room-temperature bore. The quartz sample holder was placed in a separate cryostat which was mounted in the room-temperature bore of the magnet cryostat. The monitoring and control of the sample temperature were performed using a Lakeshore 336 temperature controller. Temperatures below 4.2~K were obtained by pumping the liquid helium bath in the sample cryostat. The angular dependence was measured by rotating the magnet around the sample cryostat. This limits the angular span of the measurement since the magnet cryostat is connected to the helium exhaust line. The measured  signal reached maximal amplitude available for magnetometer already in $\mu_0 H\lesssim 2$~T, so torque in higher field (2-5~T) was measured using a Quantum Design Physical Property Measurement System - PPMS. The $\beta$-TeVO$_{4}$ crystal was mounted on a torque chip in two different orientations, and the chip on the sample board was mounted on the PPMS horizontal single-axis rotator allowing the rotation of the sample in constant magnetic fields. For the PPMS measurements the contribution to the torque signal due to gravity and puck had to be subtracted. In both experiments the magnetic field was rotated in the $a^*b$ and the $ac$ crystallographic planes. The mass of the sample was $m=1.533$~mg.
\section{Results}\label{sec:results}
\subsection{Angular dependence of magnetic torque}\label{sec:exptorque}
\begin{figure*}[tb]
	\centering
		\includegraphics[width=1.0\textwidth]{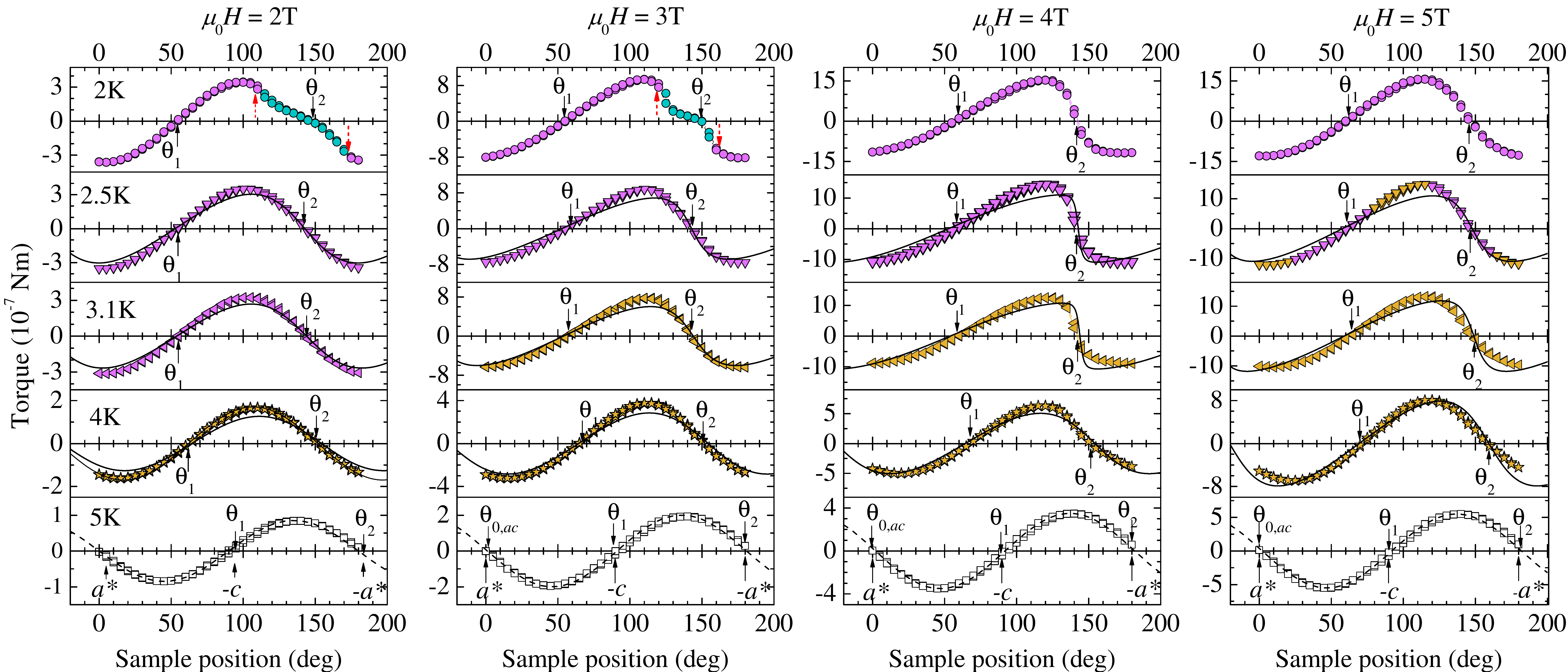}
	\caption{Angular dependencies of the torque in the $ac$ plane measured for temperatures between $2\leq T\leq 5$~K in magnetic fields $\mu_0 H=2$, 3, 4 and 5~T. Different magnetic phases are mapped by different colors based on our analysis presented in Sec. \ref{sec:phasediagram}: PM - white, SDW - gold, spin-stripe - magenta and VC - cyan. Dashed lines represent fit to \eqref{eq:torqueac} while solid lines represent the torque obtained in case of a simple model of in-plane reorientation of spins (see Sec.\ \ref{sec:spinreor}). Solid black arrows denote $\theta_1$ and $\theta_2$ defined in text. The VC phase is established in the range of angles between the dashed red arrows.}
	\label{fig2}
\end{figure*}
\indent The angular dependence of magnetic torque measured in $\mu_0 H=2, 3, 4$ and 5~T at selected temperatures between $2 \leq T \leq 5$~K is shown in Figs.\ \ref{fig2} and \ref{fig3} for the $ac$ and the $a^*b$ plane, respectively. Full set of measurements is given in Supplementary material \cite{Herak-suppl}. In case of a linear response of magnetization to magnetic field the angular dependence of the magnetic torque can be described by the following expressions \cite{Pregelj-2016}
\begin{subequations}\label{eq:torque}
\begin{align}\label{eq:torqueasb}
	\tau_{a^*b}&= \dfrac{m}{2M_{mol}} H^2 \: \Delta\chi_{a^*b} \: \sin(2\theta -2\theta_{0,a^*b} ),\\
\label{eq:torqueac}
	\tau_{ac}&= \dfrac{m}{2M_{mol}} H^2 \:\Delta\chi_{ac} \:\sin(2\theta-2\theta_{0,ac}),
\end{align}
\end{subequations}
where $m$ is the mass of the sample and $M_{mol}$ is the molar mass. $\Delta\chi_{a^*b}$ and $\Delta \chi_{ac}$ are the susceptibility anisotropies, i.e., the difference between the value of maximal and minimal susceptibility components in the $a^*b$ and $ac$ planes, respectively. For our choice of coordinate system, $\theta_{0,a^*b}$ and $\theta_{0,ac}$ represent the angles at which the torque is zero when measured from the $a^*$ axes in the $a^*b$ and $ac$ planes, respectively. For the monoclinic \bto the symmetry requires that the crystal $b$ axis is one of the magnetic eigenaxes. Consequently, in \eqref{eq:torqueasb} $\theta_{0,a^*b}=0$ and $\Delta \chi_{a^*b} = \chi_{a^*} - \chi_b$. In the $ac$ plane directions of the eigenaxes are not restricted, so $\theta_{0,ac}$ is likely to be finite \cite{Newnham}. From Eqs.\ \eref{eq:torqueasb} and \eref{eq:torqueac} it follows that another zero of torque curves can be found at angles $\theta_{0,a^*b}\pm 90^{\circ}$ and $\theta_{0,ac}\pm 90^{\circ}$ for the $a^*b$ and the $ac$ plane, respectively. Here we also define the two other zeroes we observe in torque curves in the $ac$ plane, $\theta_1$ and $\theta_2$ (\fref{fig2}), which correspond to the directions of magnetic eigenaxes in the $ac$ plane. In the paramagnetic state $\theta_1=\theta_{0,ac}+90^{\circ}$, $\theta_2=\theta_{0,ac}+180^{\circ}$. However, in the magnetically ordered phases this will not always hold, and $\theta_1$ and $\theta_2$ are more generally defined as the zeros around which $(d\tau/d \theta)_{\theta\rightarrow\theta_1}>0$ and $(d\tau/d \theta)_{\theta\rightarrow\theta_2}<0$, respectively.\\ 
\begin{figure*}[tb]
	\centering
		\includegraphics[width=1.00\textwidth]{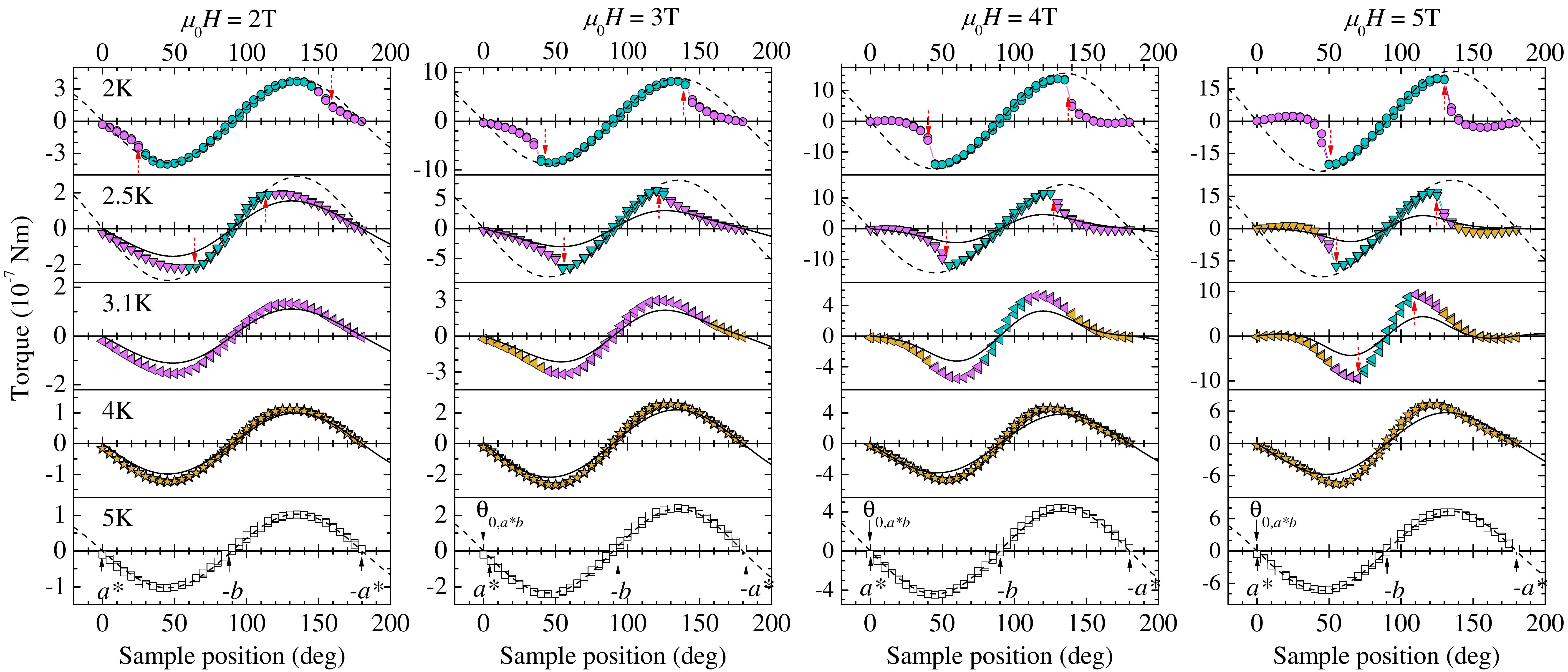}
	\caption{Angular dependencies of the torque in the $a^*b$ plane measured for temperatures between $2\leq T\leq 5$~K in magnetic fields $\mu_0 H=2$, 3, 4 and 5~T. Different magnetic phases are mapped by different colors based on our analysis presented in Sec. \ref{sec:phasediagram}: PM - white, SDW - gold, spin-stripe - magenta and VC - cyan. Dashed lines represent fit to \eqref{eq:torqueasb} while solid lines represent the torque obtained in case of a simple model of in-plane reorientation of spins (see Sec.\ \ref{sec:spinreor}). The dashed black lines for $\mu_0 H>2$~T were obtained by multiplying the line for $\mu_0 H=2$~T by $H^2/2^2$ since, according to \eqref{eq:torqueasb} the torque is expected to have a $H^2$ dependence. The VC phase is established in the range of angles between the dashed red arrows.}
	\label{fig3}
\end{figure*}
\indent At low magnetic fields, i.e., $\mu_0 H \leq 1$ T, the measured torque is well described by Eqs.\ \eref{eq:torque} down to the lowest accessible temperatures \cite{Pregelj-2016,Herak-suppl}. In the paramagnetic (PM) state, i.e., at $T = 5$ K, this is even valid in higher fields, as all the data collected at this temperature  can be described by Eqs.\ \eref{eq:torque}, yielding $\theta_{0,a^*b}=0$ and $\theta_{0,ac}$=0. This signifies that, within the experimental uncertainty, the crystal axes $a^*$, $b$ and $c$ are also the magnetic eigenaxes  of the PM state, as reported previously from the low-field torque measurements \cite{Pregelj-2016}. Indeed, the fit (black dashed lines in Figs.\ \ref{fig2} and \ref{fig3} for $T=5$~K) yields the values of susceptibility anisotropies at 5~K: $\Delta\chi_{ac}=\chi_{a^*}-\chi_{c} = -7\cdot 10^{-4}$ and $\Delta \chi_{a^*b}=\chi_{a^*}-\chi_{b} = -8\cdot 10^{-4}$, in good agreement with the previous report \cite{Pregelj-2016}.\\
\indent As the temperature is lowered, the angular dependence of the torque curves measured for $\mu_0 H\geq 2$~T deviates from the simple expressions \eref{eq:torque} in both probed planes, as can be seen in Figs.\ \ref{fig2} and \ref{fig3}. There are two possible reasons for this deviation: (1) the phase diagram of \bto is highly anisotropic \cite{Savina-2015, Weickert-2016,Pregelj-2019PRB}, so we are probing different magnetic phases at different temperatures, magnitudes and directions of the magnetic field, or (2) the rotation of constant magnetic field at constant temperature is changing the magnetic structure within the particular magnetic phase (spin reorientation phenomenon). Regarding the feature (1), here we jump ahead and color the data points in Figs.\ \ref{fig2} and \ref{fig3} according to the magnetic phase of the system for the specific $(\theta, T, H)$. The phases are attributed according to our results described in more detail in Sec.\ \ref{sec:phasediagram}.\\
\indent For the torque measured in the $ac$ plane we notice (\fref{fig2}) that the angles $\theta_1$ and $\theta_2$ shift with temperature. These shifts are related to the rotation of the magnetic eigenaxes with temperature, previously observed in the low-field torque experiments \cite{Pregelj-2016,Herak-suppl}. In \sfref{fig4}{(a)} we plot the temperature dependencies of $\theta_1$ and $\theta_2$ in the $ac$ plane. The change is the largest in the temperature range $3.8 $~K $ \leq T \leq 5$~K and then it saturates with $\theta_1$ and $\theta_2$ at the angles which are in the vicinity of the $[10\overline{1}]$ and $[\overline{1}0\overline{1}]^*$ crystal axes, respectively. For $\mu_0 H=2$ and 3~T there is a small increase observed in $\theta_2$ at $T \approx 2.2$~K marked by the arrow in \sfref{fig4}{(a)} at $T_{N_3}$ which is absent for $\mu_0 H=4$ and 5~T. This small change in $\theta_2$ was noticed previously in the torque measurements in low field \cite{Pregelj-2016} and it signifies a phase transition form the spin-stripe phase to the VC phase, which does not occur for $\mu_0 H\geq 4$~T in the $ac$ plane \cite{Pregelj-2015,Savina-2015,Weickert-2016,Pregelj-2019PRB}. Interestingly, this increase is absent for $\theta_1$ for all applied fields (\fref{fig4}) signifying that around the corresponding magnetic eigenaxes at $\theta_1$ the system remains in the spin-stripe phase for all $\mu_0 H \geq2$~T. Looking back at \fref{fig2} we see a deformation of the torque curves for $\mu_0 H \leq 3$~T at $T= 2$~K in the vicinity of the $\theta_2$ which represents the phase transition between the spin-stripe and the VC phase induced only by the \emph{rotation of a constant magnetic field at a constant temperature}. The VC phase in the $ac$ plane (\fref{fig2}) is observed only in the range of angles between the dashed red arrows around $\theta_2$.\\
\indent There are, however, two peculiar results in the torque data which were not observed previously in the low-field measurements. Firstly, for $\mu_0 H \geq 3$~T we find that $\theta_2 - \theta_1 \neq 90^{\circ}$, as can be seen in \sfref{fig4}{(b)}. Secondly, the magnitude of the positive and negative amplitude $\tau_0^{+}$ and $\left| \tau_0^{-} \right|$ is different. In \sfref{fig4}{(c)} we plot the ratio $\tau_0^{+} /\left| \tau_0^{-} \right|$ which significantly differs from 1 for $\mu_0 H \geq 3$~T. Interestingly, both effects are most pronounced for $\mu_0 H=4$~T. We will discuss these results in Sec.\ \ref{sec:disc} after we present the full analysis of our data in sections \ref{sec:spinreor} and \ref{sec:phasediagram}. \\
\indent In the $a^*b$ plane, however, the zeroes of the torque curves do not change with temperature (\fref{fig3}), which corroborates the fact that the $b$ axis is indeed one of the magnetic eigenaxis. The angular dependence of the torque curves in the $a^*b$ plane differs significantly from the curves measured in the $ac$ plane. Specifically, for $T \lesssim 3$~K, a kink, denoted by dashed red arrows, is observed at specific angles around the $b$ axis. This kink represents a phase boundary which we attribute to the boundary between the VC phase and the spin-stripe phase, based on the phase diagram \cite{Savina-2015,Weickert-2016,Pregelj-2019PRB}, as well as our result described in Sec.\ \ref{sec:phasediagram}. In particular, the torque in the VC phase is well described by expression \eref{eq:torqueasb} plotted by the dashed black lines for $T \leq 2.5$~K while at angles and temperatures which probe the other two phases the torque magnitude significantly reduces at angles around $\pm a^*$ direction. This is in agreement with the previous low-field results, which showed that the anisotropy in the $a^*b$ plane is much larger in the VC phase than in other two phases \cite{Pregelj-2016,Herak-suppl}.\\
\begin{figure}[tb]
	\centering
		\includegraphics[width=1.00\columnwidth]{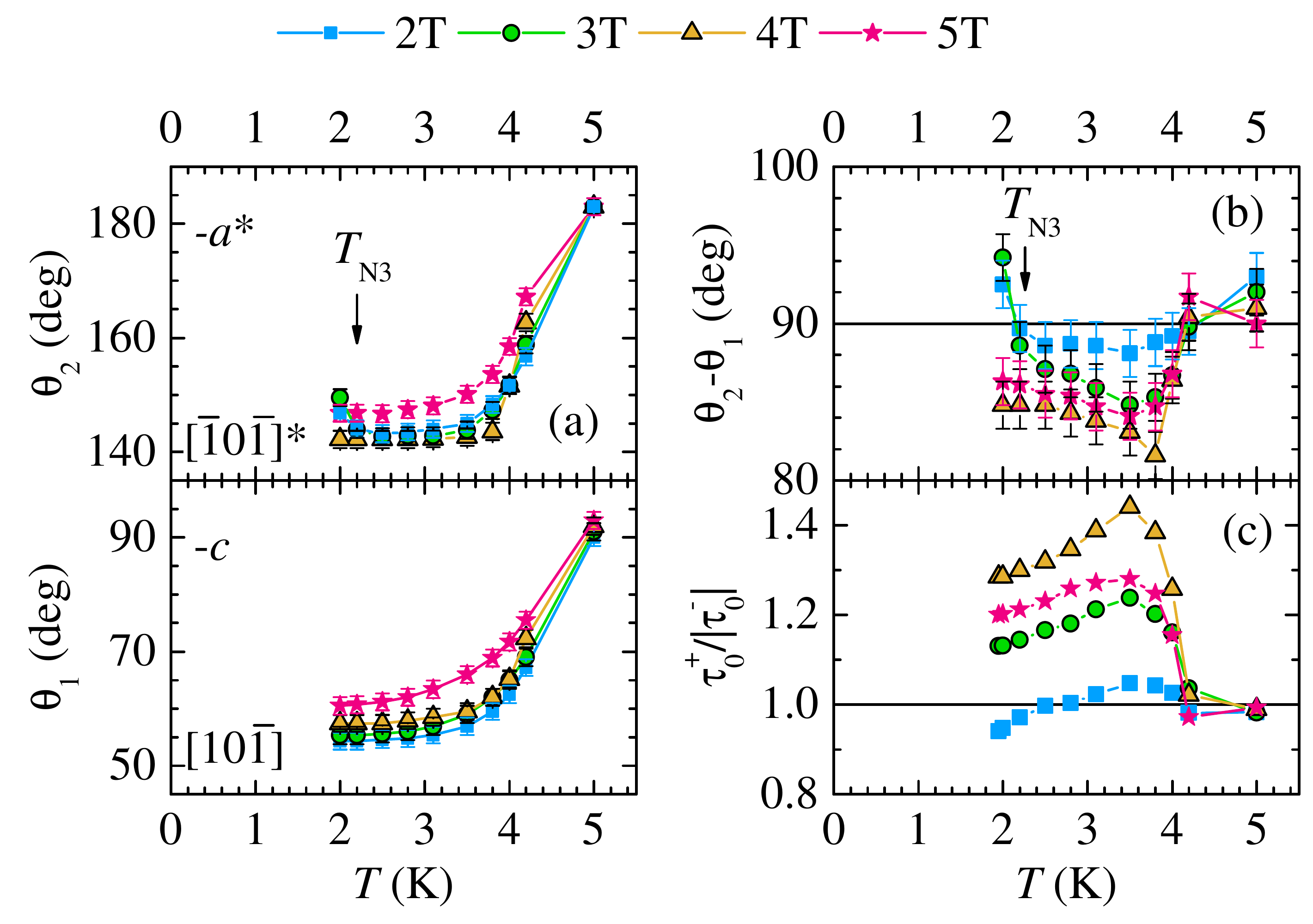}
	\caption{(a) Temperature dependence of $\theta_1$ and $\theta_2$ obtained in the $ac$ plane. $-c$ axis is found at 90$^{\circ}$, and $-a^*$ axis at 180$^{\circ}$, nominally. $\theta_1$ and $\theta_2$ are defined in \fref{fig2}. (b) Temperature dependence of the difference $\theta_2-\theta_1$ in the $ac$ plane. Black arrows in (a) and (b) denote the phase transition at $T_{N_3}$ observed only for $\mu_0 H=2$ and 3~T. (c) Temperature dependence of the ratio of the absolute values of the positive and negative torque amplitudes, $\tau_0^{+} /\left| \tau_0^{-} \right|$ obtained in the $ac$ plane.}
	\label{fig4}
\end{figure}
\indent While the phase boundary between the VC and the spin-stripe phase is easily observed in the torque curves (Figs.\ \ref{fig2} and \ref{fig3}), the distinction between the SDW and spin-stripe phase is not that obvious. In fact, in the temperature range which should span both of these phases \cite{Savina-2015, Weickert-2016,Pregelj-2019PRB}, the shape of the torque curves seems unchanged. Therefore, we needed to analyze our curves in more detail, described in Sec.\ \ref{sec:phasediagram}, to obtain the phase boundaries which allowed us to color the data points in Figs.\ \ref{fig2} and \ref{fig3} according to the magnetic phases of the system for each specific $(\theta, T, H)$. The similarity of torque curves in the SDW and the spin-stripe phases suggests that, from the point of view of macroscopic anisotropy, there is no significant difference between those two phases. This is consistent with the magnetic structure of these two phases, since the component of the magnetic moment in the spin-stripe phase, which accompanies the main SDW component \cite{Pregelj-2016}, is apparently too small to be detected in our macroscopic measurements. Also, the spin-stripe phase is dynamic \cite{Pregelj-2019} which may average out the additional anisotropies to those of the parent SDW.\\
\indent The angular dependence of the torque curves shown in Figs.\ \ref{fig2} and \ref{fig3} is characteristic for an antiferromagnetic material with the magnetic-field-induced spin reorientation \cite{Herak-2013,Novosel-2019}. We will use this fact to interpret the measured curves within the simple spin reorientation model in Sec.\ \ref{sec:spinreor}.
\subsection{Macroscopic spin reorientation model}\label{sec:spinreor}
\indent In this section we show that the spin reorientation takes place in the SDW and spin-stripe phases of \btos. For that purpose we employ a very simple macroscopic spin reorientation model to calculate the angular dependence of the torque curves and then compare them to the experimental curves. The obtained field-induced spin reorientation is then added as a new element to the anisotropic phase diagram of \bto in Sec.\ \ref{sec:phasediagram}.\\
\indent We start by noting that the angular dependence of the torque in the SDW and the spin-stripe phases in \fref{fig2} and \ref{fig3} resembles the angular dependence of torque of an easy-axis antiferromagnet measured in magnetic field comparable to the spin-flop field \cite{Herak-2013,Novosel-2019}, as we will elaborate below. In the SDW phase the moments on an individual chain are collinear. Yet there are two magnetically inequivalent chains, which differ in the magnitude and the orientation of the magnetic moments that point along the $a$ axis on one chain and along the $c$ axis on the other \cite{Pregelj-2016}. From the macroscopic point of view, this effectively results in the magnetic anisotropy that is otherwise characteristic of an uniaxial antiferromagnetic system (see \fref{fig7} in Sec.\ \ref{sec:disc}). Indeed, in both the SDW and the spin-stripe phase, the eigenvalues of the magnetic susceptibility tensor have a temperature dependence characteristic for the easy-axis antiferromagnet \cite{Blundell}: one component reduces with temperature in a manner similar to easy axis susceptibility, and two other components are almost temperature independent and very similar in magnitude \cite{Pregelj-2016}. We thus asign the easy axis direction to the direction of the smallest component of the susceptibility tensor which we term $\chi_{\|}$. This allows us to employ a simple model of spin-axis reorientation to simulate the measured torque curves in \btos.\\
\indent It is well known that when magnetic field in easy-axis antiferromagnets is applied along the easy axis direction, the spin-flop phenomenon is observed at the characteristic field called the spin-flop field $H_{SF}$ \cite{Neel-1952,Blundell}. This happens due to the competition between the magnetocrystalline anisotropy energy and Zeeman energy. In case of uniaxial anisotropy energy we have for the spin-flop field $H_{SF}=\sqrt{2\:K/(\chi_{\perp}-\chi_{\|})}$, where $K$ is the magnetocrystalline anisotropy energy constant and $\chi_{\|}$ and $\chi_{\perp}$ are susceptibility components along the easy and hard axis, respectively \cite{Neel-1952,Yosida-1951}.\\
\indent When the magnetic field  is applied at some angle $\psi \neq 90^{\circ}$ with respect to the easy axis direction, the spin axis (spins) will reorient in such a way to minimize the total energy \cite{Neel-1952,Yosida-1951}. This reorientation influences the angular dependence of the torque curves, which then deviates from the expressions like \eqref{eq:torque}. Moreover, the angular dependence of the torque curves will be different for different planes of rotation of magnetic field since the corresponding spin reorientations will be different. Indeed, the torque curves measured for $T<5$~K in the $ac$ plane resemble the torque curves measured in uniaxial antiferromagnet in the plane containing the easy axis for the case when the spin reorientation occurs in the plane of measurement \cite{Novosel-2019}. For that particular case there is an analytical expression for the angle $\theta_0$ that the spin axis makes with the direction of the easy axis in the applied magnetic field $H$ is \cite{Yosida-1951}
\begin{equation}\label{eq:spinaxis}
	\theta_0 (H,\psi) = \dfrac{\sin 2\psi}{\cos 2\psi - H^2/H_{SF}^2}.
\end{equation}
This reorientation is pronounced already for $H$ significantly lower than $H_{SF}$ \cite{Herak-2013} and it clearly reflects as a change of the angular dependence of the measured torque curves with respect to the low-field result in \eqref{eq:torque}\cite{Herak-2013,Novosel-2019}, and as observed in \fref{fig2}.\\
\indent We thus proceed by writing the standard susceptibility tensor of the easy-axis antiferromagnet as
	\begin{equation}\label{eq:tensorSusc}
	\mathbf{\hat{\chi}} = 
	\begin{bmatrix}
\chi_{\|} & 0 & 0\\
	0 & \chi_{IM} & 0 \\
0& 0 & \chi_{\perp}
	\end{bmatrix},
	\end{equation}
where $\chi_{\|}$, $\chi_{IM}$ and $\chi_{\perp}$ are easy, intermediate and hard axis, respectively. For simplicity, we expressed the tensor \eref{eq:tensorSusc} in the coordinate system spanned by its eigenaxes (magnetic eigenaxes). In this simple model, the reorientation of the spin-axis from the easy axis direction in an applied magnetic field can be described by the rotated susceptibility tensor
\begin{equation}\label{eq:chi}
	\hat{\boldsymbol{\chi}}_{rot}(\theta_0) = \mathbf{R}(\theta_0) \cdot \hat{\boldsymbol{\chi}} \cdot \mathbf{R}^T(\theta_0),
\end{equation}
where $\theta_0$ is given by the expression \eref{eq:spinaxis} and $\mathbf{R}$ is the rotation matrix. \\
\indent To calculate the angular dependence of magnetic torque we first need to derive the spin axis reorientation in applied magnetic field. Using the previously measured susceptibility tensor at low fields \cite{Pregelj-2016}, we can determine the magnetization $\mathbf{M}=\hat{\boldsymbol{\chi}}_{rot}(\theta_0)\cdot \mathbf{H}$, where $\hat{\boldsymbol{\chi}}_{rot}(\theta_0)$ is given by expression \eref{eq:chi}. This is justified by the fact that the eigenaxes rotation with temperature [\fref{fig4}{(a)}] is practically equivalent to the low-field result \cite{Pregelj-2016,Herak-suppl}). The rotation of the spin axis in the applied magnetic field is then simulated by the rotated susceptibility tensor \cite{Yosida-1951,Novosel-2019} given by expression \eref{eq:chi}. We further simplify the model by assuming that the spin axis can rotate in applied magnetic field only in the $ac$ plane, so the rotation matrix in \eref{eq:chi} is given by  
	\begin{equation}\label{eq:rotMatrix}
	\mathbf{R}(\theta_0) = 
	\begin{bmatrix}
	\cos(\theta_0) & 0 & \sin(\theta_0)\\
	0 & 1 & 0 \\
	-\sin(\theta_0) & 0 & \cos(\theta_0)
	\end{bmatrix},
	\end{equation}
where the angle $\theta_0(H,\psi)$ is given by expression \eref{eq:spinaxis}. This choice of rotation is explained above.\\
\indent The magnetic torque is calculated from the expression $\boldsymbol{\tau}= m/ M_{mol}\;\mathbf{M} \times \mathbf{H}$, where the factor $m/ M_{mol}$ ($m$ is mass and $M_{mol}$ is molar mass) is introduced because the magnetic susceptibility is measured in units of emu/mol. The only free parameter in our simulation is the value of the spin-flop field $H_{SF}$. For torque curves measured at different temperatures a different value of $H_{SF}$ was presumed in order to obtain the best fit to the measured curves. This is in accordance with a simple expression for the spin-flop field of uniaxial antiferromagnets, $H_{SF} \propto\sqrt{ K/\Delta \chi}$, where both $K$ and $\Delta \chi=\chi_{\perp}-\chi_{\|}$ are temperature-dependent. However, considering very low temperatures of these phases as well as the fact that all the phases are antiferromagnetic,  the magnetocrystalline anisotropy constant $K$ can be assumed to be temperature-independent. The values of the spin-flop field used for each temperature are given in Table \ref{tabHsf}. The spin-flop field $H_{SF}$ increases as the temperature increases and is largest in the vicinity of $T_{N_1}$ which is reasonable since $\Delta \chi (T)$ significantly reduces on heating as $T\rightarrow T_{N_1}$ \cite{Pregelj-2016}.\\
\begin{table}[tb]
	\centering
		\caption{Values of the spin-flop field $H_{SF}$ used in spin-reorientation model in Sec.\ III~B.}
		\begin{tabular*} {\columnwidth}{@{\extracolsep{\fill} }c|cccccc}
		\hline \hline
			$T$ (K) & 4.0 & 3.8 & 3.5 & 3.1 & 2.8 & 2.5\\\hline
			$\mu_0 H_{SF}$ (T) & 8.0 & 4.3 & 4.2 & 4.2 & 3.9 & 3.8\\
			\hline \hline			
		\end{tabular*}
	\label{tabHsf}
\end{table}
\indent The calculated torque curves obtained for the rotation of magnetic field in the $ac$ plane are shown as solid lines in \fref{fig2}. The shape of the curves, as well as the absolute magnitude of the torque reproduce  the experimental data quite well for the entire temperature and field range except at angles for which the system is in the VC phase (between the red dashed arrows for $\mu_0 H=2$~T and 3~T), thus verifying that the spin reorientation indeed takes place in \btos. The corresponding angular dependencies of the spin-axis directions with respect to the easy-axis direction, $\theta_0(H,\psi)$, calculated using \eqref{eq:spinaxis} and $H_{SF}$ values from Table \ref{tabHsf}, can be found in Supplementary material \cite{Herak-suppl}. The calculated torque curves have slightly smaller amplitude than the measured curves (\fref{fig2}), but this difference is well within the boundaries produced by the uncertainty of the measured tensor components. For $T\leq 2.2$~K, the calculations were not performed since in low fields at these temperatures the system is in the VC phase \cite{Pregelj-2016}. \\ 
\indent The rotation in the $a^*b$ plane was simulated by assuming again that the spins can rotate only in the $ac$ plane [rotation matrix \eref{eq:rotMatrix}] while using the same values for $H_{SF}$ given in Table \ref{tabHsf}. In this case only the component of the magnetic field along the $a^*$ axis contributes to rotation. This simple model captures the experimental result for the $a^*b$ plane rather well (\fref{fig3}) where results of simulation are plotted by solid lines (see also Supplementary material \cite{Herak-suppl}), except for the the range of angles at which the VC state is established (between the dashed red arrows). This result further corroborates the finding that the field-induced spin reorientation takes place in the SDW and spin-stripe phases of \btos. The calculated $\theta_0(H,\psi)$, used in the simulation of the torque in the $a^*b$ plane can be found in Supplementary material \cite{Herak-suppl}.\\
\indent The apparent reduction of the torque amplitude around the $\pm a^*$ axes observed in the $a^*b$ plane at low temperatures in $\mu_0 H\geq 4$~T, as compared to the lower field, can now be easily understood. When the magnetic field is in the vicinity of the $\pm a^*$ axes it significantly rotates the spin axis which moves further away from the $\pm a^*$ axes toward the $c$ axis, which, consequently reduces the susceptibility anisotropy in the $a^*b$ plane.\\
\indent Finally, we note that the torque curves in the VC state can be described by the expression \eref{eq:torqueasb}, which is plotted by the dashed black lines in \fref{fig3}. This suggests that in the VC phase there is no spin reorientation, at least not in the range of magnetic fields applied in our experiment.
\subsection{\texorpdfstring{Anisotropic phase diagram of $\bm{\beta-\mathrm{TeVO}_4}$}{Anisotropic phase diagram of beta-TeVO4}}\label{sec:phasediagram}
\begin{figure}[tb]
	\centering
		\includegraphics[width=\columnwidth]{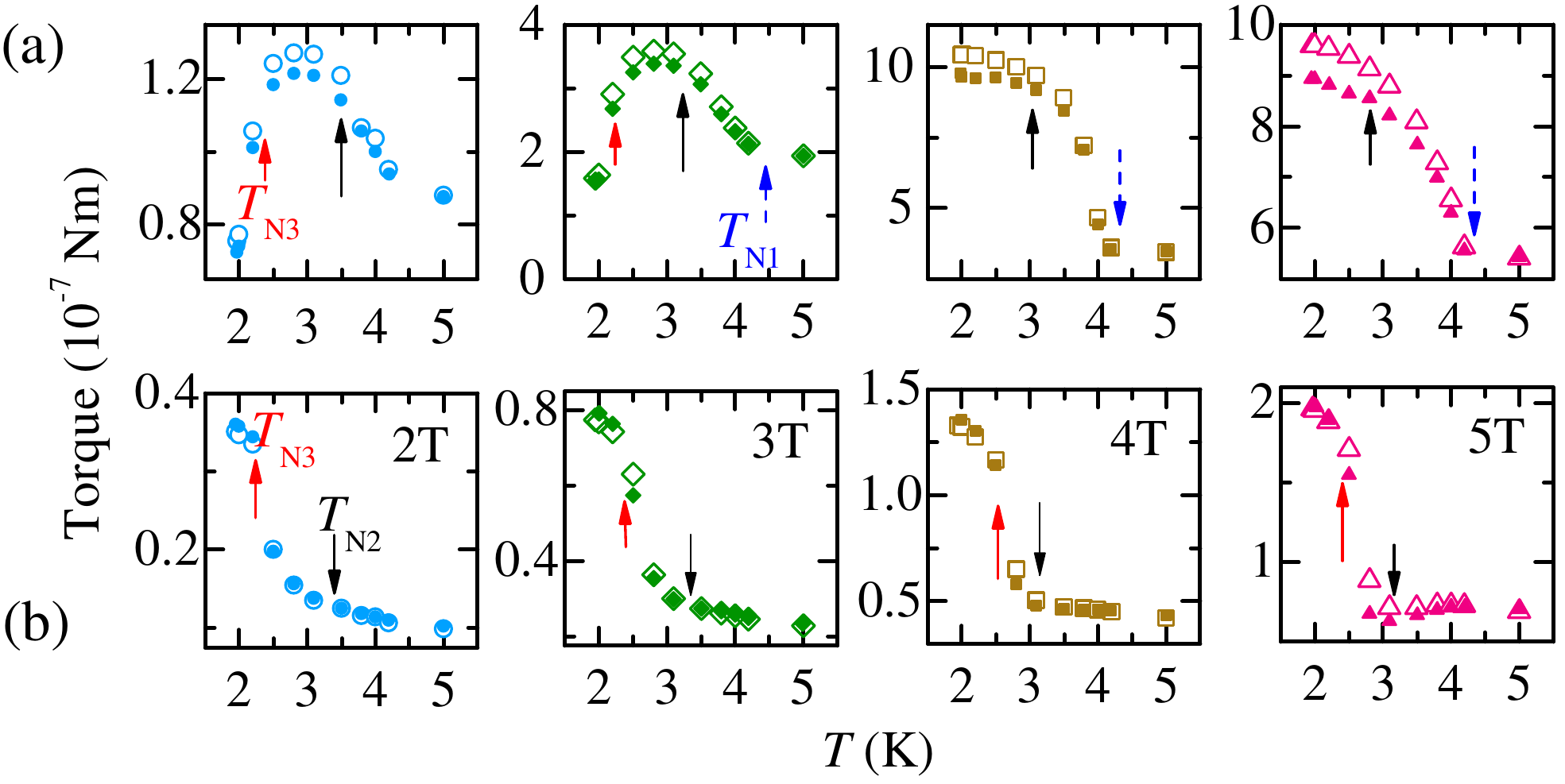}
	\caption{Temperature dependence of torque amplitude measured in (a) the $ac$ plane for $135^{\circ}$ and (b) the $a^*b$ plane for sample position of $125^{\circ}$ (\fref{fig3}) and. The phase transition temperatures are denoted by dashed blue ($T_{N_1}$, solid black $T_{N_2}$ and solid red ($T_{N_3}$) arrows. Empty and solid symbols denote the data obtained, respectively, from torque curves recorded while rotating the sample from $0^{\circ}$ to $180^{\circ}$ and back.}
	\label{fig5}
\end{figure}

\indent The anisotropic phase diagram of \bto is obtained from the torque curves measured at different temperatures and magnetic fields in the $ac$ and $a^*b$ planes shown (Figs.\ \ref{fig2} and \ref{fig3}, including a full set of measurements given in Supplementary material \cite{Herak-suppl}). From the measured data we extract the temperature-dependent torque curves for each angle at a constant field and field-dependent torque curves for each angle at constant temperature. As a characteristic example we show in \fref{fig5} the temperature-dependent torque data obtained for the specific sample orientations in different magnetic fields for both the $a^*b$ and $ac$ plane (more examples of obtained curves are given in Supplementary material \cite{Herak-suppl}). The angular-dependent anisotropic phase diagram obtained from the torque measurements has been reported recently for zig-zag spin chain system linarite \cite{Feng-2018}. In that experiment magnetic field was swept at constant temperature and specific direction, which allowed for the detailed mapping of the many phases of linarite. Angular dependencies, however, were not recorded, so the possible field-induced reorientations might have been missed in that system. While our finite set of $(T,H)$ values results in somewhat larger uncertainty of the obtained phase boundaries, it has an obvious advantage of vividly revealing an occurrence of the so far undetected spin reorientations in \bto (Figs.\ \ref{fig2} and \ref{fig3}).\\
\indent The phase transition temperature $T_{N_1}\approx 4.6$~K from the SDW to the spin-stripe phase does not change significantly with the field magnitude or direction \cite{Savina-2015,Weickert-2016,Pregelj-2019PRB}. We established from our curves measured at $T=5$~K and $T=4.2$~K in both planes, that at the former temperature the system is in the PM state [curves can be fitted to Eqs.\ \eref{eq:torque} for all $\mu_0H\geq 2$~T], while at $T=4.2$~K the data can no longer be described by Eqs.\eref{eq:torque} (see Supplementary material \cite{Herak-suppl}). Therefore, we set $T_{N_1}=4.6$~K with the error bar reflecting the uncertainty in the assigned value. However, for $\mu_0 H\geq 4$~T the temperature reduces to $T_{N_1}\approx 4.2$~K in the vicinity of $\theta_2$.\\
\indent The phase transition temperature $T_{N_2}$ is difficult to distinguish because it does not involve a sharp change of torque.  Still, we are able to determine $T_{N_2}$, with some uncertainty, by using our low-field results where we established that in the $ac$ plane the torque increases rapidly on cooling below $T_{N_1}$ and then almost saturates at $T_{N_2}$, while in the $a^*b$ plane it changes weakly on cooling below $T_{N_1}$ and then starts to increase more rapidly at $T_{N_2}$,  (\fref{fig5} and Supplementary material \cite{Herak-suppl}). At $T_{N_3}$ a sharp change is observed (jump in the $ac$ plane and kink in the $a^*b$ plane, both followed by a weaker change of torque magnitude on further cooling, \fref{fig5}). More detail can be found in Supplementary material \cite{Herak-suppl}.\\ 
\begin{figure*}[tb]
	\centering
				\includegraphics[width=0.8\textwidth]{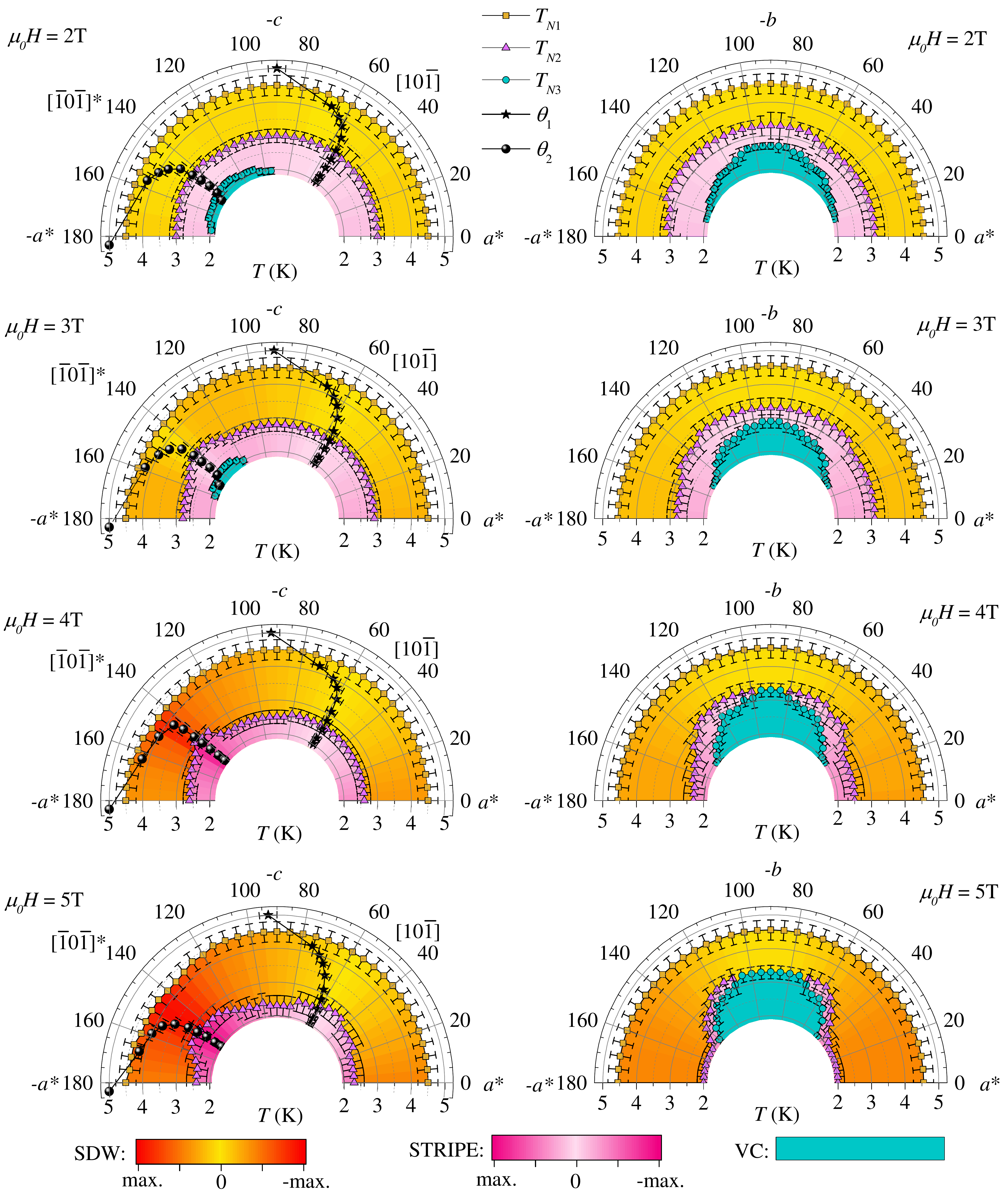}
	\caption{Anisotropic phase diagram of \bto in the $ac$ (left) and $a^*b$ (right) planes measured in magnetic fields $\mu_0 H=2-5$~T. Temperature change of the position of two magnetic eigenaxes, at angles $\theta_1$ and $\theta_2$ in the $ac$ plane, is also shown. The three magnetic phases, SDW, spin-stripe and spiral, are designated according to literature \cite{Pregelj-2015,Pregelj-2016,Pregelj-2018,Pregelj-2019PRB}. In the SDW phase, magnetic axes rotate and settle in the vicinity of the $\left[\overline{1}0\overline{1}\right]^*$, $\left[10\overline{1}\right]$ and $b$ crystallographic directions in the VC phase. In the SDW and spin-stripe phase the spin reorientation occurs when magnetic field $\mu_0 H\geq 2$~T is applied. The color bar represents the calculated $\theta_0$ - the angle of the rotation of spin axes from the easy-axis direction, which is used in the simulation of the measured magnetic torque curves as described in Sec.\ \ref{sec:spinreor}. The values ''max'' and ''-max'' represent the maximal angle $\theta_0$ measured from the easy axis direction at specific temperature. The value of $\theta_0$ for specific temperatures and fields is plotted in Figs.\ S6 and S7 in the Supplementary materials \cite{Herak-suppl}. At lower temperatures ''max'' and ''-max'' equals $+90^{\circ}$ and $-90^{\circ}$, which are equivalent values for antiferromaget. In the VC phase there is no spin reorientation in the applied magnetic field.}
	\label{fig6}
\end{figure*}
\indent The resulting anisotropic phase diagram is shown in \fref{fig6} where error bars reflect the uncertainties in assigning the values to different transition temperatures. The colors reflect different magnetic phases, as well as the spin reorientation obtained from our model in Sec. \ref{sec:spinreor} for the SDW and the spin-stripe phases. Unlike the previously reported phase diagrams which were measured by applying magnetic field along the crystal $a$, $b$ and $c$ axes, the phase diagram in \fref{fig6} contains an additional information on  what happens to the different phase boundaries as the magnetic field rotates in the chosen crystal planes. This information is used to color the data points of measured torque curves in Figs.\ \ref{fig2} and \ref{fig3}.\\
\indent The phase transition at $T_{N_1}$ from the paramagnetic to the SDW state appears almost isotropic in both planes. In the $ac$ plane, however, we observe the decrease of $T_{N_1}$ for $\mu_0 H\geq 4$~T when the field is applied along the $\theta_2$ angle, which corresponds to the macroscopic easy axis direction. This effect complies with the usual response of an antiferromagnet in high magnetic fields, which decreases the N\'{e}el temperature.\\
\indent The phase boundary between the SDW and the spin-stripe phase, $T_{N_2}$, is, on the other hand, highly anisotropic (\fref{fig6}). Magnetic field suppresses $T_{N_2}$ in the $ac$ plane and does not introduce additional magnetic anisotropy. In the $a^*b$ plane $T_{N_2}$ is suppressed anisotropically with the largest decrease observed for $H \| \pm a^*$ axis, while around the $\pm b$ $T_{N_2}$ appears less sensitive to $H$.\\
\indent The phase boundary at $T_{N_3}$ representing the phase transition to the VC phase is very anisotropic in both planes (\fref{fig6}). In the $ac$ plane $T_{N_3}$ is observed, in the given temperature range, only around the easy axis at $\theta_2$ for $\mu_0 H\leq 3$~T. Applying the magnetic field in the $ac$ plane thus suppresses the VC phase to lower temperatures. On the contrary, in the $a^*b$ plane, $T_{N_3}$ increases as the field increases and for $H$ close to the $b$-axis reaches a value of $T_{N_3}\approx 3.3(2)$~K. Furthermore, for $\mu_0 H<4$~T we find that the phase boundary $T_{N_3}$ corresponds to transition from the spin-stripe phase to the VC phase, while for $\mu_0 H\geq 4$~T, $T_{N_3}$ represents the phase boundary between the SDW and the VC phases (see the phase diagram for $a^*b$ plane in \fref{fig6}).\\
\indent All these results are in excellent agreement with the literature while adding a novel insight into the behavior of the phase boundaries as the magnetic field rotates. In particular, this work adds to the anisotropic phase diagram of \bto the insight into the magnetic-field-induced spin reorientation, expressed in \fref{fig6} by a different color scale for the SDW (yellow-red) and the spin-stripe phase (pale pink - bright magenta). The color in our phase diagram corresponds to the calculated value of $\theta_0(H,\psi)$ obtained from the simple model presented in Sec.\ \ref{sec:spinreor} (the calculated values of the angle $\theta_0$ can be found in Supplementary material \cite{Herak-suppl}). We find that the spin reorientation is the largest for $\mu_0 H\geq 4$~T in the $ac$ plane around the easy axis direction $\theta_2$. For $\mu_0 H \leq 3$~T there is a modest spin reorientation when the field is applied between the two axes $\theta_1$ and $\theta_2$, but not when it's applied along those axes. In the $a^*b$ plane a relatively strong reorientation appears when the field is in the vicinity of the $a^*$ axes. This reorientation increases with the applied field, but it does not reach maximal values as it does for the $ac$ plane because the easy axis is not in the plane of rotation.
\section{Discussion}\label{sec:disc}
\begin{figure}[tb]
	\centering
		\includegraphics[width=\columnwidth]{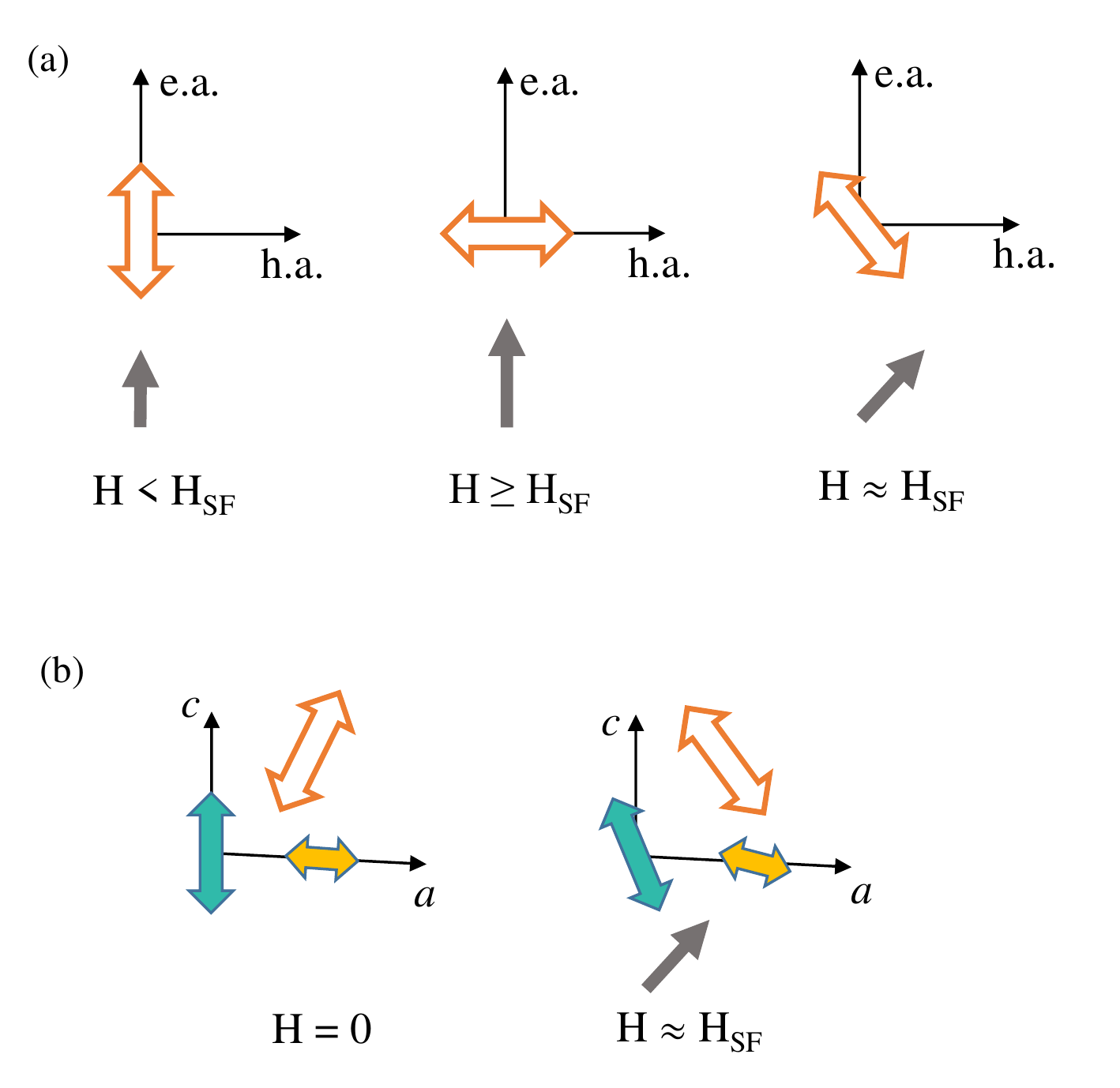}
	\caption{(a) Spin axis reorientation in an easy-axis antiferromagnet. Orange empty double arrow represents the spin axis orientation. Left: When a field $H<H_{SF}$ is applied along the easy axis (e.a.), the spin axis remains along the easy axis direction. Middle: When $H\geq H_{SF}$ is applied along the easy axis, the spin axis reorients along the hard axis (h.a.). Right: When the magnetic field comparable to the spin-flop field is applied in some general direction, the spin axis reorients to minimize the total energy. (b) In the SDW phase of \bto there are two spin density waves with magnetic moments oriented along the $\pm c$ and $\pm a$ directions, denoted by blue and yellow double arrows. The reorientation of the macroscopic spin axis (orange empty double arrow) in applied field is a consequence of the coherent  reorientation of the magnetic moments on individual chains, while introducing the change in angle between the moments on different chains.}
	\label{fig7}
\end{figure}

\indent The magnetic phase diagram of \bto is highly complex \cite{Pregelj-2015,Savina-2015,Weickert-2016,Pregelj-2019PRB}. Taking into account all the previous results as well as the results obtained in this work (\fref{fig6}) we establish three important features of the magnetic phases of \btos: (1) the anisotropic phase diagram of \bto is strongly dependent on the temperature as well as on the magnitude and the direction of the applied magnetic field, (2) in the SDW phase we observe a large rotation of magnetic eigenaxes with temperature and (3) in the SDW and the spin-stripe phase the magnetic field induces a spin reorientation which is also dependent on the field direction. Here we emphasize that the rotation of magnetic eigenaxes with temperature [feature (2)], observed in the entire range of fields up to 5~T [\sfref{fig4}{(a)} and Refs.\ \onlinecite{Pregelj-2016} and \onlinecite{Herak-suppl}] should not be confused with the spin reorientation we discuss in Sec.\ \ref{sec:spinreor} [feature (3)] which occurs at a constant temperature and is induced by magnetic fields $\mu_0 H\geq 2$~T. While feature (1) is a direct consequence of the exchange anisotropies which are present in \btos, features (2) and (3) are more surprising. We next discuss what lies behind them.\\
\indent In \bto the rotation of magnetic eigenaxes occurs for the larger part in the SDW phase. Combining this with the fact that magnetic order in the SDW phase is characterized by two magnetic chains, one having magnetic moments aligned along the $a$ axis and the other along the $c$ axis \cite{Pregelj-2016}, suggests that the temperature-dependent rotation of the magnetic eigenaxes [feature (2)] can be related to the different rate of increase of magnetic moments on different chains on cooling. This scenario is supported by the different magnitude of the magnetic moments extracted from the neutron diffraction data for the different chains in the zero field at $T=3.5$~K, $\mu_c/\mu_a = 0.7/0.4$ \cite{Pregelj-2016}. Furthermore, if magnetic moments for both chains were equal, the measured anisotropy in the $ac$ plane would be almost zero, and macroscopically, we would observe an easy plane and not an easy axis behavior. While the almost perpendicular orientation of the spins on two inequivalent chains \cite{Pregelj-2016} most likely reflects the significance of interchain Dzyaloshinskii-Moriya interaction (DMI), the different values of magnetic moments on two inequivalent chains point to a further complexity of interactions in \btos.\\
\indent In order to explain the magnetic-field-induced reorientation of the magnetic moments within the SDW and spin-stripe phases [feature (3)], we employ in Sec.\ \ref{sec:spinreor} a simple model of spin reorientations developed for uniaxial antiferromagnets with easy-axis anisotropy to simulate the measured torque curves. The resulting agreement between the measured and the simulated curves allows us to characterize the field-induced changes in magnetic structure as spin reorientations, while the value of the critical magnetic field $H_{SF}$ is also derived from the model (Table \ref{tabHsf}). The observed spin reorientation is driven by the competition between the different anisotropies which are responsible for the zero-field spin orientation and the Zeeman energy. Our simple macroscopic model would suggest uniaxial anisotropy with the easy axis direction corresponding to $\theta_2$ i.e. crystal direction $\approx <1\:0\:1>^*$ for $T\lesssim 4$~K. However, we should be careful not to draw a parallel between this simple macroscopic result and the microscopic anisotropic interactions which are present in \btos. While our simple result could be mapped to a spin Hamiltonian with only symmetric anisotropic exchange, the anisotropic Hamiltonian of \bto allows for both symmetric, as well as antisymmetric (Dzyaloshinskii-Moriya) anisotropic interactions. The latter was discussed qualitatively in Ref. \onlinecite{Weickert-2016}, while the former was used to theoretically study the magnetic phase diagram of \bto and was able to reproduce many of the features, despite the simplicity of the model \cite{Singhania-2020}. The complexity of the magnetic phases in \bto suggest that all these anisotropies of both intra- and interchain interaction play a significant role in this system.\\
\indent We stress here that the magnetic structures of the SDW and the spin-stripe phases are significantly more complicated than the simple collinear uniaxial antiferromagnet. In that context our spin-reorientation model  is too simple and it cannot capture the different microscopic anisotropic interactions needed to describe all the phases of \btos. Regardless, the observed torque results can offer qualitative insight into the microscopic behavior in applied magnetic field. In order to visualize this, we first present a sketch of the spin reorientation in the textbook example of an uniaxial antiferromagnet in \sfref{fig7}{(a)}. We plot the orientation of the spin axis (empty solid double-arrow) with respect to easy axis for different magnitudes and orientations of magnetic field. While in the uniaxial easy-axis antiferromagnet the experimentally derived orientation of the spin axis actually corresponds to the orientation of spins, in \bto we deal with a more abstract easy axis. Here the derived easy axis corresponds to the direction along which the susceptibility tensor has a minimal value (direction corresponding to angle $\theta_2$ in Figs.\ \ref{fig2} and \ref{fig6}, which is neither along the $a$ nor the $c$ axis, but somewhere in-between). In the SDW phase of \bto two inequivalent chains are sketched in \sfref{fig7}{(b)} by solid blue and yellow double arrows signifying two SDWs \cite{Pregelj-2016}. Obviously, the rotation of the macroscopic easy axis (empty orange double-arrow) in \bto reflects the rotation of the actual spins on chains.\\
\indent In the applied magnetic field the spins in \bto can either rotate coherently (by the same angle on both chains), or in some more complicated manner. For the coherent rotation we would expect no difference between the simple uniaxial antiferromagnet and our system. However, there are two experimental features which disagree with such simple case. Firstly, in the experiment we find the difference $\theta_2-\theta_1\neq 90^{\circ}$ [\sfref{fig4}{(b)}], while the model suggests $\theta_2-\theta_1=90^{\circ}$. Secondly, the ratio of the positive and negative torque amplitude $\tau_0^{+}/|\tau_0^{-}|\neq 1$ in the experiment [\sfref{fig4}{(c)}], while equal amplitudes are predicted by the model. These differences allow us to propose a likely scenario of actual spin reorientation in \btos: The spins on each chain rotate in the applied magnetic field, but the angle between the two perpendicular SDWs on neighboring chains changes in a manner that depends on the field magnitude and direction. Since the moments on different chains are almost, but not exactly perpendicular, and also have different magnitudes, this type of reorientation can produce the observed difference $\theta_2 - \theta_1 \neq 90^{\circ}$ as well as different ratio of positive and negative torque amplitudes, shown in Figs.\ \sref{fig4}{(b)} and \sref{fig4}{(c)}. Such reorientation scenario in \bto is sketched in \sfref{fig7}{(b)}. Future refinements of magnetic structure in applied magnetic field should test this scenario.\\
\indent Finally, we note that the VC phase seems to be remarkably robust in the applied magnetic field, as can be seen from \fref{fig3} where the torque in the VC phase follows the expression \eref{eq:torqueasb}. This signifies that there are no spin reorientations in the VC phase, at least not for the investigated values of temperature and magnetic field.
\section{Conclusion}\label{sec:concl}
\indent The detailed magnetic torque study shows that applying the magnetic field in the magnetically ordered phases of \bto does not only change the phase boundaries between the competing phases, but also introduces the spin reorientation phenomena in the SDW and the spin-stripe phases. We use a simple spin-reorientation model, originally developed to describe the spin reorientation of uniaxial antiferromagnet with an easy-axis anisotropy \cite{Neel-1952,Yosida-1951}, to simulate the measured torque curves. This allows us to add a new important element to the anisotropic $(T,\mathbf{H})$ phase diagram of this system, namely spin-axis orientation in the applied magnetic field. The spin reorientation occurs because of the competition between the Zeeman energy and magnetic anisotropy energy. While our analysis does not capture the complexity of the anisotropic interactions in \btos, which should include both symmetric and antisymmetric (DMI) anisotropic exchange, we were able to propose a more realistic model of spin reorientation [\sfref{fig7}{(b)}] which could account for the observed asymmetry of the torque curves. Our results suggest the spin reorientation is constrained to the $ac$ crystal plane. The proposed scenario should be tested by the future microscopic measurements. The magnitude of exchange anisotropies in \bto must be comparable to the Zeeman energy corresponding to the spin-reorientation field $\mu_0 H_{SF}\approx 4$~T obtained at the lowest temperatures. From the macroscopic point of view, there is no significant difference between the reorientation in the SDW and the spin-stripe phases.\\
\indent Our results also reveal that the phase diagram is anisotropic in the $ac$ plane. The fact that the phase boundaries are equivalent for $H\| a$ and $H\| c$  is merely a coincidence that occurs because the magnetic eigenaxes are rotated by $\approx 45^{\circ}$ from the crystal axes. This result emphasizes the importance of studying the anisotropic phase diagrams in the low-dimensional frustrated spin systems by experiments which utilize magnetic field rotation, such as torque magnetometry, rather than just by measurements with the magnetic field applied only in the directions of the main crystal axes.\\
\indent Spin reorientations at the microscopic level should be revealed in future microscopic studies of magnetic structures in \bto in the applied magnetic field by advanced techniques, such as neutron diffraction and nuclear magnetic resonance (NMR). The experimentally determined phase diagram of \bto and the spin reorientation presented in this work present an important step forward in establishing the model of anisotropic magnetic interactions which are responsible for the formation of exotic magnetic phases in this and similar frustrated quantum spin systems.
\begin{acknowledgments}
M.~H., N.~N. and M.~D. acknowledge full support of their work by the Croatian Science Foundation under Grant No. UIP-2014-09-9775. M.~H. and M.~D. acknowledge support by COGITO project ''Theoretical and experimental investigations of magnetic and multiferroic materials'' co-funded by the Croatian Ministry of Science and Education and The French Agency for the promotion of higher education, international student services, and international mobility. D.~A. acknowledges financial support from the Slovenian Research Agency (Core Research Funding No. P1-0125 and Projects No. J1-9145 and No. N1-0052). M.~H. is grateful to Dominik Cin\v{c}i\'{c} from the Department of Chemistry at the Faculty of Science of the University of Zagreb for checking the crystal axes directions of the sample used in measurements and to Tomislav Ivek from Institute of Physics in Zagreb for critically reading the manuscript. Crystal structure shown in \fref{fig1} was drawn using 3D visualization program VESTA \cite{Vesta}.
\end{acknowledgments}
\bibliographystyle{apsrev4-1}
\bibliography{Ref}
\end{document}